\documentclass[12pt]{article}
\usepackage[top=25mm,bottom=25mm,left=25mm,right=25mm]{geometry}
\usepackage{authblk}

\usepackage{graphicx}
\usepackage{amsmath}
\usepackage{amssymb}
\usepackage{amsfonts}
\usepackage{mathtools}
\usepackage{derivative}

\title{Nonlinear dynamic elastic moduli from equilibrium stress fluctuations}
\author[1]{F.E. Garbuzov}
\author[1]{Y.M. Beltukov}
\affil[1]{Ioffe Institute, 26, Polytekhnicheskaya, St.Petersburg, 194021, Russia}
\date{}

\newcommand{\dd}[1]{\mathop{\mathrm{d}#1}} % differential
\newcommand{\CBK}{C^\text{BK}}
\newcommand{\CBKc}[1]{C^\text{BK}_{#1}}
\newcommand{\NBK}{N^\text{BK}}
\newcommand{\NBKc}[1]{N^\text{BK}_{#1}}
\newcommand{\Dlin}{\Delta H_\Gamma^\text{lin}}
\newcommand{\Dnl}{\Delta H_\Gamma^\text{nl}}

\newcommand{\C}{C^\text{dyn}}
\newcommand{\N}{N^\text{dyn}}

\newcommand{\avg}[1]{\left\langle #1 \right\rangle}
\newcommand{\avgs}[1]{\langle #1 \rangle}

\newcommand{\avgb}[1]{\bigl\langle #1 \bigr\rangle}

\begin{document}

\maketitle

\begin{abstract}
Fluctuation formulas for elastic and viscoelastic moduli allow their computation from equilibrium molecular dynamics simulations, avoiding explicit nonequilibrium deformation protocols. While such expressions are well established for the quasi-static moduli, and also the linear dynamic moduli, no fluctuation formula exists for the nonlinear time-dependent moduli that govern anharmonic viscoelastic response under finite time-dependent strains. In this work we derive transient-time correlation function expressions for both the linear and the nonlinear dynamic moduli, starting from the DOLLS/SLLOD equations of motion for irrotational motion. The resulting formulas involve equilibrium time correlations of the stress tensor and Born–kinetic terms, and they recover the known quasi-static and linear dynamic results in the appropriate limits.
\end{abstract}

\section{Introduction}

Knowing the elastic and viscoelastic moduli of a material is essential for predicting its mechanical response to external loads. Molecular dynamics offers a way to compute these moduli directly from the underlying interatomic interactions, thereby linking macroscopic mechanical properties to the microscopic system description. 

To identify the elastic moduli, one can either perform nonequilibrium simulations and deduce the moduli directly from the stress response to an applied strain, or perform equilibrium simulations and extract them from the stress or strain fluctuations. While the equilibrium approach is computationally attractive, its application requires rigorous fluctuation formulas.

For the linear quasi-static (zero-frequency) elastic constants, fluctuation expressions based on strain fluctuations~\cite{ParrinelloRahmanStrainFluctuations1982} and stress fluctuations~\cite{RayMoodyMolecularDynamics1985,LutskoGeneralizedExpressions1989,SquireHoltIsothermalElastic1969,BavaudChoquardStatisticalMechanics1986} were obtained more than four decades ago. The extension of fluctuation methods to quasi-static nonlinear elastic moduli followed more recently~\cite{KarmakarLernerAthermalNonlinear2010}. For time-dependent deformations, a fluctuation formula for the linear viscoelastic moduli is also available~\cite{WilliamsEvansViscoelasticProperties2009}.

A gap nevertheless remains in the understanding of the nonlinear dynamic mechanical response: no fluctuation formula exists for the nonlinear time-dependent moduli that govern the leading anharmonic contributions in viscoelastic materials subjected to small but finite time-dependent strains. The present work aims to fill this gap.

The paper is organized as follows. Section~\ref{sec:microscopic} introduces the microscopic equations of motion suitable for modeling an induced strain and gives the microscopic expression for the stress tensor. Section~\ref{sec:dynamic_moduli} contains the main derivation: we obtain the nonequilibrium probability distribution function in Sec.~\ref{sec:f_neq} and use it in Sec.~\ref{sec:ttcf} to evaluate the averaged stress in the regime of small but finite strains, arriving at explicit expressions for the linear and nonlinear dynamic moduli. The paper concludes with a summary and outlook in Sec.~\ref{sec:outro}.

\section{Microscopic equations}
\label{sec:microscopic}

The dynamics of a system of interacting particles is described by the Hamiltonian
\begin{equation} \label{hamilt}
    H (\{\vec r_a\}, \{\vec p_a\}) = \sum_{a=1}^N \frac{\vec p_a \cdot \vec p_a}{2m_a} + U(\{\vec r_a\}),
\end{equation}
where $\vec r_a$, $\vec p_a$, and $m_a$ are the coordinates, momenta, and masses of particle $a$, respectively. The phase space point is denoted by $(\{\vec r_a\}, \{\vec p_a\}) = (\vec r_1, \dots, \vec r_N, \vec p_1, \dots \vec p_N)$, $U$ is the interaction potential, and the central dot denotes the scalar (inner) product.

The equations of motion derived from this Hamiltonian describe a closed system at equilibrium. To model nonequilibrium behavior corresponding to deformation, these equations must be modified. %In that case, the Hamiltonian in Eq.~\eqref{hamilt} is interpreted as the internal energy of the system.

\subsection{Nonequilibrium equations of motion}
Several formulations of the equations of motion for homogeneous nonequilibrium flows have been developed, and there is ongoing discussion in the literature regarding their validity.  

An early Hamiltonian approach is the DOLLS method \cite{HooverEvansLennardJonesTriplepoint1980},
where the total Hamiltonian in Eq.~\eqref{hamilt} is augmented by a term that couples the particle momenta to the streaming velocity field:
\begin{equation} \label{hamilt_dolls}
    H_\text{DOLLS} = H + \sum_a \vec p_a \vec r_a : L,
\end{equation}
with the streaming velocity gradient tensor $L = \nabla \vec v^{\,T}$. The dyadic (outer) product is implied by the juxtaposition of two vectors, and $:$ denotes double contraction, so that the second term reads $\sum_a \sum_{ij} p_{a,i} r_{a,j} L_{ij}$ in index notation. The corresponding Hamiltonian equations of motion are
\begin{align}
\label{dolls_r}
    \dot{\vec r}_a &= \frac{\vec p_a}{m_a} + L\cdot\vec r_a,\\
\label{dolls_p}
    \dot{\vec p}_a &= -\pdv{U}{\vec r_a} - L^T\cdot\vec p_a.
\end{align}

It was soon discovered that the DOLLS method produces erroneous normal stresses in the nonlinear regime when simulating planar shear~\cite{EvansMorrissNonlinearresponseTheory1984,EvansMorrissNonNewtonianMolecular1984}. This issue was addressed by the SLLOD algorithm, which replaces the transpose $L^T$ by $L$ in the momentum equation:
\begin{align}
    \dot{\vec r}_a &= \frac{\vec p_a}{m_a} + L\!\cdot\!\vec r_a, \\
    \dot{\vec p}_a &= -\pdv{U}{\vec r_a} - L\!\cdot\!\vec p_a.
\end{align}
%Although SLLOD is non‑Hamiltonian, it is exact for adiabatic flows and forms the basis of many practical NEMD simulations. 
Although these equations are not Hamiltonian, they preserve phase space volume~\cite{EvansMorrissStatisticalMechanics2008}.

The SLLOD equations have been widely used in nonequilibrium molecular dynamics, particularly for planar shear flow~\cite{EvansMorrissNumberDependence1989, SandbergHeyesSelfDiffusion1995, MundySiepmannDecaneUnder1995, AoyagiDoiMolecularDynamics2000, FadaeiBjorlingUnravelingPressure2023}.
For elongational and more general flows the proper‑SLLOD (p‑SLLOD) algorithm~was introduced~\cite{BaigEdwardsProperApproach2005}, which however raised a number of objections~\cite{DaivisToddSimpleDirect2006}. Moreover, p-SLLOD produces the stress tensor that differs from the standard one that follows from the Irving–Kirkwood procedure~\cite{IrvingKirkwoodStatisticalMechanical1950}. For these reasons we do not adopt p‑SLLOD in the present work.

DOLLS and SLLOD differ only when the velocity‑gradient tensor $L$ is non‑symmetric, i.e., when the flow involves
rotation (non‑zero vorticity). Because our focus is on elastic
response, we restrict ourselves to irrotational deformations, where
\(L\) is symmetric (\(L = L^T\)).  Under this condition the DOLLS and SLLOD
equations coincide identically.  Therefore, throughout this paper we
employ the following DOLLS/SLLOD equations of motion:
\begin{align}
	\dot{\vec r}_a &= \frac{\vec p_a}{m_a} + L(t)\!\cdot\!\vec r_a, \\
	\dot{\vec p}_a &= -\pdv{U}{\vec r_a} - L^T(t)\!\cdot\!\vec p_a,
	\qquad L = L^T.
\end{align}
Although we assume $L$ to be symmetric in what follows, we retain $L^T$ so that our derivations remain applicable in the DOLLS framework even without the symmetry condition on $L$.

\subsection{Pulled-back variables}
In solids, deformation is described by the deformation gradient $F$, mapping reference positions $\vec R_a$ to current positions $\vec r_a$:
\begin{equation} \label{r_canon}
    \vec r_a = F(t) \cdot \vec R_a.
\end{equation}
The canonically conjugate momenta $\vec P_a$ corresponding to $\vec R_a$ are defined as
\begin{equation} \label{p_canon}
    \vec p_a = F^{-T}(t) \cdot \vec P_a,
\end{equation}
and the velocity gradient is related to $F$ by
\begin{equation} \label{L}
    L(t) = \dot F(t) \cdot F^{-1}(t).
\end{equation}
We refer to $\vec R_a$ and $\vec P_a$ as the pulled-back coordinates and momenta. 

In the literature, it has been a common practice to apply the canonical transformation defined by Eqs.~\eqref{r_canon} and \eqref{p_canon} to the Hamiltonian~\eqref{hamilt} to derive fluctuation formulas for quasi-static (zero-frequency) linear~\cite{ParrinelloRahmanCrystalStructure1980, BavaudChoquardStatisticalMechanics1986, LutskoGeneralizedExpressions1989} and nonlinear elastic moduli~\cite{KarmakarLernerAthermalNonlinear2010}.
In the dynamic case, the DOLLS/SLLOD equations~\eqref{dolls_r}--\eqref{dolls_p} are consistent with this transformation in the sense that, for a suddenly applied deformation,
\begin{equation}
    F(t) = \begin{cases}
        I, &t<0,\\ F, &t>0,
    \end{cases} \qquad
    \dot F(t) = (F - I) \delta(t),
\end{equation}
the pulled-back variables $\vec R_a$ and $\vec P_a$ remain continuous at $t=0$.
This implies that the laboratory coordinates and momenta must undergo the jumps
\begin{align}
    \vec r_a(0+) &= F \cdot \vec r_a(0-),
\label{r_transform}\\
    \vec p_a(0+) &= F^{-T} \cdot \vec p_a(0-).
\label{p_transform}
\end{align}

We note that relations~\eqref{r_transform} and~\eqref{p_transform} hold in both DOLLS and SLLOD frameworks under the irrotational flow assumption adopted here. However, when rotational motion is present (i.e., when $L$ is nonsymmetric), the two approaches diverge: in SLLOD the pulled-back momenta suffer a jump determined by the skew-symmetric part of $L$, and the jump relation for the laboratory momenta cannot be expressed as simply as in Eq.~\eqref{p_transform}. Since we restrict ourselves to irrotational motion, this distinction lies outside our scope.

By virtue of Eqs.~\eqref{r_transform} and \eqref{p_transform}, a suddenly applied strain produces an instantaneous change in the laboratory coordinates and momenta; thus, in this sense, they are strain-dependent. The pulled-back variables, in contrast, are free of such jumps: their strain dependence enters only through their time evolution. Expressing the DOLLS/SLLOD equations~\eqref{dolls_r} and \eqref{dolls_p} in terms of these pulled-back variables yields
\begin{align}
    \dot{\vec R}_a &= C^{-1} \cdot \frac{\vec P_a}{m_a}, \hspace{15mm} C = F^T\cdot F,\\
    \dot{\vec P}_a &= -\pdv{U(\{F\cdot \vec R_a\})}{\vec R_a},
\end{align}
where $C$ is the right Cauchy--Green deformation tensor.

\subsection{Strain}

For small but finite strains, we use the Lagrangian finite strain tensor
\begin{equation}
    E = \frac12 (C - I) = \frac12 (F^T \cdot F - I).
\end{equation}
Due to the pullback transformation in Eqs.~\eqref{r_canon} and \eqref{p_canon}, the Hamiltonian $H$ becomes an explicit function of the strain:
\begin{equation} \label{hamilt_pb}
    H(\Gamma, E) = \sum_a \frac{\vec P_a \cdot C^{-1} \cdot \vec P_a}{2m_a} + U(\{F\cdot \vec R_a\}),
\end{equation}
where $\Gamma = (\{\vec R_a\}, \{\vec P_a\})$ denotes the set of all pulled-back particle coordinates and momenta, and we assume that $U$ depends on $F$ only through $C$ (and therefore through $E$). The simplest example of such a potential is a pair potential which depends only on intermolecular distances $(\vec r_a - \vec r_b)^2 = \vec r_{ab}^2 = \vec R_{ab} \cdot C \cdot \vec R_{ab}$, thus $C$ plays the role of a metric tensor for the deformed space~\cite{ParrinelloRahmanCrystalStructure1980}. However, we keep $U$ in unspecified form for generality.

The stress conjugate to the Lagrangian strain is the second Piola–Kirchhoff stress tensor:
\begin{equation} \label{pk2_der}
    S(\Gamma, E) = \frac1{V_0} \pdv{H}{E} = \frac1{V_0} \sum_a \biggl(-C^{-1} \cdot \frac{\vec P_a \vec P_a}{m_a} \cdot C^{-1} + C^{-1} \cdot \pdv{U}{\vec R_a} \vec R_a \biggr)
    %\frac1{2V_0} \left(F^{-1} \cdot \pdv{H}{F} + \pdv{H}{F^T} \cdot F^{-T}\right),
\end{equation}
where $V_0$ is the reference (undeformed) volume.
Using the usual expression for the second Piola-Kirchhoff stress through the Cauchy stress $\sigma$
\begin{equation}
    S = (\det F) F^{-1} \cdot \sigma \cdot F^{-T},
\end{equation}
and the canonical transformation (Eqs.~\eqref{r_canon}, \eqref{p_canon}) we recover the expression for the virial stress:
\begin{equation} \label{sigma}
    \sigma = \frac1V \sum_a \left(\pdv{U}{\vec r_a} \vec r_a - \frac{\vec p_a \vec p_a}{m_a} \right), \quad V = V_0 \det F,
\end{equation}
thus making it a microscopic expression for the Cauchy stress. We note that the virial stress is not symmetric in general, but under our assumptions it is symmetric, as we will discuss later.

The rate of change of internal energy under deformation is
\begin{equation}
    \dot H = V \sigma : L = V_0 S : \dot E,
\end{equation}
which takes this form in both SLLOD and DOLLS approaches even without the assumptions of irrotational motion~\cite{EvansMorrissStatisticalMechanics2008}.
Thus, the total change of internal energy is given by
\begin{equation}\label{delta_H}
    \Delta H_\text{tot}(t) = V_0 \int_0^t S(t') : \dot E(t') \dd {t'},
\end{equation}
where $S(t) = S(\Gamma(t), E(t))$ is time-dependent through the phase space point $\Gamma(t)$ and the strain $E(t)$.

\subsection{Tensor symmetries}
We assume the potential $U$ to be arbitrary, subject only to the constraints of translational and rotational invariance. These symmetries imply the conservation of linear and angular momentum, which requires that the total force and total torque in the system vanish:
\begin{equation}\label{eq:pot_sym}
    \sum_a \pdv{U}{\vec r_a} = \vec{0}, \quad \sum_a \vec r_a \times \pdv{U}{\vec r_a} = \vec{0}. %\quad \text{where} \quad \vec f_a = -\pdv{U}{\vec r_a}.
\end{equation}
The rotational symmetry condition can be equivalently formulated as the requirement that the following tensor is symmetric:
\begin{equation} \label{eq:rot_sym}
    \sum_a \vec r_{a} \pdv{U}{\vec r_a} = \sum_a \pdv{U}{\vec r_a} \vec r_{a},
\end{equation}
from which the symmetry of the virial (Cauchy) and of the second Piola-Kirchhoff stresses follows: 
\begin{equation}
    \sigma = \sigma^T, \quad S = S^T.
\end{equation}
%Here and throughout this paper, Latin subscripts $a$, $b$, and $c$ denote particle indices, while $i,j,k,l,$ etc. denote Cartesian components.

A violation of the symmetry condition in Eq.~\eqref{eq:rot_sym} indicates the presence of internal torques and couple stresses. Such systems are described in the continuum limit, e.g. by Cosserat (or micropolar) elasticity. By enforcing Eq.~\eqref{eq:rot_sym}, we restrict our analysis to systems whose macroscopic behaviour is governed by classical Cauchy elasticity.

\section{Dynamic elastic moduli}
\label{sec:dynamic_moduli}

Suppose we consider a system of particles representing a (visco-)elastic material. Upon deformation, we seek an expression for the stress as a functional power series in the strain history~\cite{GarbuzovBeltukovGeneralizationNonlinear2024, PipkinSmallFinite1964, FindleyLaiCreepRelaxation1976}:
\begin{align} \label{S_multiint}
    \avgs{S(t)}_\text{neq} =& \int_0^t \C(t-t_1) : \dot E(t_1)\dd{t_1} \nonumber\\
    &+ \int_0^t \int_0^t \N(t-t_1, t-t_2) :: \dot E(t_1) \dot E(t_2)\dd{t_1}\dd{t_2} + \dots,
\end{align}
where $\C$ is the fourth-order tensor of linear dynamic (time-dependent) viscoelastic moduli, and $\N$ is the sixth-order tensor of nonlinear moduli.

\subsection{Nonequilibrium distribution function}
\label{sec:f_neq}

Deformation is a nonequilibrium process, and the average in Eq.~\eqref{S_multiint} must be taken over a nonequilibrium ensemble. We assume that prior to deformation the system is described by a canonical ensemble, so that the initial distribution function is given by
\begin{equation} \label{f_0}
    f_0(\Gamma_0) = f_\text{eq}(\Gamma_0, 0),
\end{equation}
where $\Gamma_0$ is the phase space point at the initial time $t=0$ and the canonical distribution function $f_\text{eq}$ is defined by
\begin{equation} \label{f_eq}
    f_\text{eq}(\Gamma, E) = Z^{-1}(E) e^{-\beta H(\Gamma, E)}, \quad Z(E) = \int e^{-\beta H(\Gamma, E)} \dd {\Gamma}, \quad \beta = (k_B T)^{-1}.
\end{equation}
Here, the strain-dependent Hamiltonian is given by Eq.~\eqref{hamilt_pb}, $Z$ is the partition function, $k_B$ is the Boltzmann constant, and $T$ is the temperature.

If the equations of motion preserve phase space volume (as in the DOLLS or SLLOD approaches), then the nonequilibrium distribution function $f$ at phase point $\Gamma$ at time $t$, obtained by evolving the system from $\Gamma_0$, satisfies
\begin{equation}
    f(\Gamma) = f_0(\Gamma_0).
\end{equation}
This allows us to perform the following derivations:
% \begin{align}
%     f(\Gamma) &= \frac{e^{-\beta H(\Gamma_0, 0)}}{Z(0)} = \frac{e^{-\beta H(\Gamma, E)}}{Z(0)} e^{\beta (H(\Gamma, E) - H(\Gamma_0, 0))} \nonumber\\
%     &= \frac{e^{-\beta H(\Gamma, 0)}}{Z(0)} e^{-\beta (H(\Gamma, E) - H(\Gamma, 0))} e^{\beta (H(\Gamma, E) - H(\Gamma_0, 0))} = f_0(\Gamma) e^{\beta (\Delta H_\text{tot} - \Delta H_E)},
% \label{f_neq}
% \end{align}
% where $\Delta H_\text{tot} = H(\Gamma, E) - H(\Gamma_0, 0)$ denotes the total energy change which is given by Eq.~\eqref{delta_H} and $\Delta H_E = H(\Gamma, E) - H(\Gamma, 0)$ is the energy change due to an instantaneous strain.
\begin{align}
    f(\Gamma) &= \frac{e^{-\beta H(\Gamma_0, 0)}}{Z(0)} = \frac{e^{-\beta H(\Gamma, 0)}}{Z(0)} e^{\beta (H(\Gamma, 0) - H(\Gamma_0, 0))} = f_0(\Gamma) e^{\beta\Delta H_\Gamma},
\label{f_neq}
\end{align}
where $\Delta H_\Gamma = H(\Gamma, 0) - H(\Gamma_0, 0)$ denotes the energy change due to the phase point evolution only. This energy change can be equivalently understood as the difference between the total energy change ($\Delta H_\text{tot}$) defined in Eq.~\eqref{delta_H} and the energy change due to an instantaneous strain ($\Delta H_E$):
\begin{equation} \label{Delta_H_Gamma}
    \Delta H_\Gamma = \Delta H_\text{tot} - \Delta H_E, \quad \text{where} \quad
    \begin{aligned}
        \Delta H_\text{tot} &= H(\Gamma, E) - H(\Gamma_0, 0), \\
        \Delta H_E &= H(\Gamma, E) - H(\Gamma, 0).
    \end{aligned}
\end{equation}
Using Eq.~\eqref{f_neq}, the nonequilibrium average of any phase function, including the stress tensor $S$, can be written as
% \begin{align}
%     \avg{S(t)}_\text{neq} = \int S(\Gamma) f(\Gamma, E) \dd \Gamma 
%     &= \int S(\Gamma) e^{\beta (\Delta H_\text{tot}(t) - \Delta H_E(t))} f_0(\Gamma) \dd\Gamma \nonumber\\
%     &= \avgb{S(t) e^{\beta (\Delta H_\text{tot}(t) - \Delta H_E(t))}}_0 = \avgb{S(t) e^{\beta \Delta H_\Gamma(t)}}_0,
% \label{S_avg_neq}
% \end{align}
% where $\Delta H_\Gamma = \Delta H_\text{tot} - \Delta H_E$ and brackets $\avg{}_0$ denote the average with the equilibrium unstrained distribution function $f_0$.
\begin{align}
    \avg{S(t)}_\text{neq} = \int S(\Gamma) f(\Gamma) \dd \Gamma 
    = \int S(\Gamma) e^{\beta \Delta H_\Gamma(t)} f_0(\Gamma) \dd\Gamma
    = \avgb{S(t) e^{\beta \Delta H_\Gamma(t)}}_0,
\label{S_avg_neq}
\end{align}
where brackets $\avg{}_0$ denote the average with the equilibrium unstrained distribution function~$f_0$.

\subsection{Transient-time correlation function expression}
\label{sec:ttcf}

Since our goal is to get the series expression of the form of Eq.~\eqref{S_multiint}, we expand $H(\Gamma, E)$ and $S(\Gamma, E)$ in the power series in the strain $E$ and obtain
\begin{align}
\label{S_expand}
    S(\Gamma, E) &= S(\Gamma, 0) + \pdv[delims-eval=.|]{S}{E}_{\mathrlap{E=0}} : E + \frac12 \pdv[delims-eval=.|]{S}{E,E}_{\mathrlap{E=0}} :: EE + \dots \nonumber\\
    &= \sigma + \CBK : E + \NBK :: EE + \dots,\\
\label{delta_H_E_expand}
    \Delta H_E &= H(\Gamma, E) - H(\Gamma, 0) = \pdv[delims-eval=.|]{H}{E}_{\mathrlap{E=0}} : E + \frac12 \pdv[delims-eval=.|]{H}{E,E}_{\mathrlap{E=0}} :: EE + \dots \nonumber\\
    &= V_0 (\sigma : E + \frac12 \CBK::EE + \dots),
\end{align}
where we defined the following tensors:
\begin{align}
    \CBKc{ijkl} &= \frac1{V_0}\pdv[delims-eval=.|]{H}{E_{ij},E_{kl}}_{E=0} = \pdv[delims-eval=.|]{S_{ij}}{E_{kl}}_{E=0}, 
\label{CBK_def}\\
    \NBKc{ijklmn} &= \frac1{2V_0}\pdv[delims-eval=.|]{H}{E_{ij},E_{kl},E_{mn}}_{E=0} = \frac12\pdv[delims-eval=.|]{S_{ij}}{E_{kl},E_{mn}}_{E=0},
\label{NBK_def}
\end{align}
and used the fact $\vec R_a = \vec r_a$, $\vec P_a = \vec p_a$, and $S = \sigma$ at zero strain $E=0$. 

The introduced tensors $\CBK$ and $\NBK$ correspond to the instantaneous (infinite-frequency) linear and nonlinear elastic constants~\cite{LutskoGeneralizedExpressions1989, BavaudChoquardStatisticalMechanics1986, KarmakarLernerAthermalNonlinear2010}. The superscript BK denotes “Born–kinetic,” reflecting the decomposition into potential (Born) and kinetic contributions.
Their explicit microscopic expressions are
% \begin{align}
%     \CBKc{ijkl}(\Gamma) & = \frac{1}{V_0} \left(\sum_{a,b} \pdv{U}{r_{a,i}, r_{b,k}} r_{a,j} r_{b,l} - \delta_{ik}\sum_a \left(\pdv{U}{r_{a,j}} r_{a,l} - \frac{4p_{a,j} p_{a,l}}{m_a}\right)\right),
% \label{CBK}\\
%     \NBKc{ijklmn}(\Gamma) &= \frac{1}{2} \left(\frac1{V_0}\sum_{a,b,c} \pdv{U}{r_{a,i}, r_{b,k}, r_{c,m}} r_{a,j} r_{b,l} r_{c,n} - \delta_{ik} \CBKc{jlmn} - \delta_{im} \CBKc{jnkl} - \delta_{km} \CBKc{lnij}\right),
% \label{NBK}
% \end{align}
% which are symmetrised with respect to interchange of indices $i\leftrightarrow j$, $k\leftrightarrow l$ and $m\leftrightarrow n$.
\begin{align}
&\begin{aligned}[b]
    \CBKc{ijkl} =&\
        \frac1{4V_0} \sum_{a,b} \left(\pdv{U}{r_{a,i}, r_{b,k}} r_{a,j} r_{b,l}
            + \pdv{U}{r_{a,j}, r_{b,k}} r_{a,i} r_{b,l}
            + \pdv{U}{r_{a,i}, r_{b,l}} r_{a,j} r_{b,k}
            + \pdv{U}{r_{a,j}, r_{b,l}} r_{a,i} r_{b,k} \right)\\
        &-\frac1{4V_0} \sum_a \left( \delta_{ik} \pdv{U}{r_{a,j}} r_{a,l} 
            + \delta_{jk} \pdv{U}{r_{a,i}} r_{a,l} 
            + \delta_{il} \pdv{U}{r_{a,j}} r_{a,k}
            + \delta_{jl} \pdv{U}{r_{a,i}} r_{a,k} \right) \\
        &+\frac1{V_0} \sum_a \left( \delta_{ik} \frac{p_{a,j} p_{a,l}}{m_a}
            + \delta_{jk} \frac{p_{a,i} p_{a,l}}{m_a}
            + \delta_{il} \frac{p_{a,j} p_{a,k}}{m_a}
            + \delta_{jl} \frac{p_{a,i} p_{a,k}}{m_a} \right),
\end{aligned}
\label{CBK}\\
&\begin{aligned}[b]
    \NBKc{ijklmn} = &\ \begin{aligned}[t]
        \frac{1}{16V_0} \sum_{a,b,c} \biggl(&
        \pdv{U}{r_{a,i}, r_{b,k}, r_{c,m}} r_{a,j} r_{b,l} r_{c,n}
        + \pdv{U}{r_{a,j}, r_{b,k}, r_{c,m}} r_{a,i} r_{b,l} r_{c,n} \\
        &+\pdv{U}{r_{a,i}, r_{b,l}, r_{c,m}} r_{a,j} r_{b,k} r_{c,n}
        + \pdv{U}{r_{a,j}, r_{b,l}, r_{c,m}} r_{a,i} r_{b,k} r_{c,n} \\
        &+\pdv{U}{r_{a,i}, r_{b,k}, r_{c,n}} r_{a,j} r_{b,l} r_{c,m}
        + \pdv{U}{r_{a,j}, r_{b,k}, r_{c,n}} r_{a,i} r_{b,l} r_{c,m} \\
        &+\pdv{U}{r_{a,i}, r_{b,l}, r_{c,n}} r_{a,j} r_{b,k} r_{c,m}
        + \pdv{U}{r_{a,j}, r_{b,l}, r_{c,n}} r_{a,i} r_{b,k} r_{c,m}
    \biggr) \end{aligned}\\
    &- \frac18 \bigl( \delta_{ik} \CBKc{jlmn} + \delta_{jk} \CBKc{ilmn} + \delta_{il} \CBKc{jkmn} + \delta_{jl} \CBKc{ikmn} + \delta_{im} \CBKc{jnkl} + \delta_{jm} \CBKc{inkl} \\
    &\qquad + \delta_{in} \CBKc{jmkl} + \delta_{jn} \CBKc{imkl} + \delta_{km} \CBKc{lnij} + \delta_{lm} \CBKc{knij} + \delta_{kn} \CBKc{lmij} + \delta_{ln} \CBKc{kmij} \bigr).
\end{aligned}
\label{NBK}
\end{align}

Now we can substitute the derived expansions~\eqref{S_expand} and \eqref{delta_H_E_expand} into the nonequilibrium average of the stress tensor (Eq.~\eqref{S_avg_neq}) and obtain the expression for it in the form of the multiple integral expansion as in Eq.~\eqref{S_multiint}.

First, we substitute the stress expansion~\eqref{S_expand} into the total energy change (Eq.~\eqref{delta_H}):
\begin{align}
    \Delta H_\text{tot} &= V_0 \int_0^t \bigl(\sigma(t') + \CBK(t') : E(t') + \dots\bigr) : \dot E(t') \dd {t'} \nonumber\\
    &= V_0 \int_0^t \sigma(t_1) : \dot E(t_1) \dd {t_1} + V_0 \int_0^t\int_0^{t_1} \CBK(t_1) :: \dot E(t_1)\dot E(t_2) \dd {t_1} \dd {t_2} + \dots \nonumber\\
    &= V_0 \int_0^t \sigma(t_1) : \dot E(t_1) \dd {t_1} + \frac{V_0}2 \int_0^t\int_0^{t} \CBK(\max\{t_1,t_2\}) :: \dot E(t_1)\dot E(t_2) \dd {t_1} \dd {t_2} + \dots
\label{delta_H_tot_expand}
\end{align}
Then, to simplify the derivation, we decompose $\Delta H_\Gamma$ in Eq.~\eqref{Delta_H_Gamma} into terms linear and quadratic in strain, denoted $\Dlin$ and $\Dnl$, respectively:
\begin{align}
\label{D_lin}
    \Dlin &= V_0 \int_0^t (\sigma(t_1) - \sigma(t)) : \dot E(t_1) \dd{t_1},\\
\label{D_nl}
    \Dnl &= \frac{V_0}2 \int_0^t \int_0^t \bigl(\CBK(\max\{t_1,t_2\}) - \CBK(t)\bigr) :: \dot E(t_1) \dot E(t_2) \dd{t_1} \dd{t_2}.
\end{align}
Then Eq.~\eqref{S_avg_neq} can be written as follows
\begin{equation}
\avg{S(t)}_\text{neq} = \avg{S(t) \left(1 + \beta(\Dlin + \Dnl) + \frac{\beta^2}{2}(\Dlin)^2 \right)}_0 + O(E^3).
\end{equation}
Applying the expansion of the stress tensor~\eqref{S_expand}, assuming vanishing equilibrium stress ($\avg{\sigma}_0 = 0$), and neglecting higher-order $O(E^3)$ terms, we obtain
\begin{align}
    \avg{S(t)}_\text{neq} =& \underset{\text{linear in }E}{\underbrace{\avgs{\CBK}_0 : E + \beta \avgs{\Dlin \sigma}_0}} \nonumber\\
    &+ \underset{\text{quadratic in }E}{\underbrace{\avgs{\NBK}_0 :: EE + \beta \avgs{\Dlin \CBK}_0 : E + \beta \avgs{\Dnl \sigma}_0 + \frac{\beta^2}{2}\avgs{(\Dlin)^2 \sigma}_0}}.
\end{align}
Lastly, substituting Eqs.~\eqref{D_lin} and \eqref{D_nl}, we recover the multiple-integral representation~\eqref{S_multiint}, with kernels
\begin{align}
    C_{ijkl}(t-t_1) =&\, \avgs{\CBK_{ijkl}(t)}_0 - \beta V_0 \avgs{\sigma_{ij}(t) \sigma_{kl}(t)}_0 
        + \beta V_0 \avg{\sigma_{ij}(t) \sigma_{kl}(t_1)}_0,
\label{C}\\
    N_{ijklmn}(t-t_1, t-t_2) =&\, \avgs{\NBK_{ijklmn}(t)}_0 + \frac{\beta V_0}2 \avg{\CBK_{ijkl}(t) (\sigma_{mn}(t_2) - \sigma_{mn}(t)}_0 \nonumber\\
    &+ \frac{\beta V_0}2 \avg{\CBK_{ijmn}(t) (\sigma_{kl}(t_1) - \sigma_{kl}(t)}_0\nonumber\\
    &+ \frac{\beta V_0}{2} \avg{\sigma_{ij}(t) \bigl(\CBK_{klmn}(\max\{t_1,t_2\}) - \CBK_{klmn}(t)\bigr)}_0 \nonumber\\
    &+ \frac{\beta^2 V_0^2}{2} \avgs{\sigma_{ij}(t) (\sigma_{kl}(t_1) - \sigma_{kl}(t)) (\sigma_{mn}(t_2) - \sigma_{mn}(t))}_0.
\label{N}
\end{align}
%Here, functions without arguments are evaluated at time $t$ (e.g., $\sigma = \sigma(t)$), while $\sigma(-t_1) = \sigma(t - t_1)$ and similarly for other quantities.
The derived expressions establish the time-dependence of elastic moduli in the form of time-dependent correlation functions.

We note the following symmetry properties of the derived tensors. The Born–kinetic tensors $\CBKc{ijkl}$ and $\NBKc{ijklmn}$ possess both minor symmetries ($\CBKc{ijkl} = \CBKc{jikl} = \CBKc{ijlk}$, and similarly for $\NBK$ with respect to each index pair including $mn$) and major symmetry under interchange of index pairs $\CBKc{ijkl} = \CBKc{klij}$ (and similarly for $\NBK$). In contrast, the full dynamic moduli $C_{ijkl}(t)$ and $N_{ijklmn}(t_1,t_2)$ satisfy only the minor symmetries, but not the major symmetry. Additionally, the nonlinear tensor $N_{ijklmn}(t_1,t_2)$ exhibits the symmetry $N_{ijklmn}(t_1,t_2) = N_{ijmnkl}(t_2,t_1)$.

\section{Discussion and conclusion}
\label{sec:outro}

In this paper we have derived closed-form fluctuation expressions for the linear and nonlinear dynamic elastic moduli directly from the microscopic equations of motion. Starting from the DOLLS/SLLOD dynamics under the irrotational flow assumption and employing the pulled-back representation of phase-space variables, we expanded the nonequilibrium stress average in powers of the Lagrangian strain history. This procedure yields the transient-time correlation function formulas for the linear dynamic moduli tensor $C(t)$ (Eq.~\eqref{C}) and, as the central result of this work, for the nonlinear dynamic moduli tensor $N(t_1,t_2)$ (Eq.~\eqref{N}).

The derived expressions constitute a natural extension of earlier work in several respects. In the zero-frequency limit they recover the known quasi-static fluctuation formulas for both linear~\cite{LutskoGeneralizedExpressions1989} and nonlinear~\cite{KarmakarLernerAthermalNonlinear2010} elastic constants. When nonlinearity is neglected, the linear modulus reduces to the established viscoelastic fluctuation result~\cite{WilliamsEvansViscoelasticProperties2009}. The present formalism thus unifies these previously separate results within a single coherent framework and extends them into the genuinely nonlinear dynamic regime.

From a practical standpoint, the formulas derived here enable the computation of the full time-dependent anharmonic response of materials entirely from equilibrium MD simulations, without the need to perform explicit nonequilibrium deformation runs. This is particularly advantageous for systems where nonequilibrium protocols are costly, difficult to control, or hampered by poor signal-to-noise ratios at small strains. The correlation functions appearing in Eqs.~\eqref{C} and \eqref{N} involve only equilibrium averages of stress, Born–kinetic terms, and their time-correlations, all of which are readily accessible in standard equilibrium simulations.

Several directions for future work naturally arise from our results. Extending the formalism to rotational flows, where the DOLLS and SLLOD approaches differ, and to higher-order moduli would further broaden its applicability. The method could be used to investigate the contribution of anharmonic dynamic effects in realistic models of amorphous solids, such as polymers or biological materials. Finally, a natural next step is the numerical implementation of the derived formulas and their validation against direct nonequilibrium simulations for well-characterized model systems.

%\section*{Appendix}
%todo

\bibliographystyle{ieeetr}
\bibliography{refs}

\end{document}